\newif\ifarxiv\arxivtrue

\arxivtrue     

\ifarxiv
\documentclass[12pt,a4paper]{article}

\usepackage{lmodern}
\usepackage[T1]{fontenc}
\usepackage{amsmath,amssymb,bbm} 
\usepackage{accents}
\usepackage{color}
\usepackage{graphicx}
\usepackage[nosort]{cite}
\usepackage{tensor}
\usepackage{color}

\graphicspath{{./figures/}}

\usepackage[a4paper,text={173mm,216mm},centering]{geometry}

\makeatletter \let\@keywords\@empty \let\@subject\@empty
\providecommand{\keywords}[1]{\gdef\@keywords{#1}}
\providecommand{\subject}[1]{\gdef\@subject{#1}}
\def\thetitle{\@title}
\def\theauthor{\@author}
\def\thesubject{\@subject}
\def\thedate{\@date}
\def\thekeywords{\@keywords}
\makeatother
\AtBeginDocument{%
\hypersetup{pdftitle={\thetitle}}%
\hypersetup{pdfauthor={\theauthor}}%
\hypersetup{pdfsubject={\thesubject}}%
\hypersetup{pdfkeywords={\thekeywords}}%
}

\providecommand{\hypersetup}[1]{}

\hypersetup{plainpages=false}
\hypersetup{pdfpagemode=UseNone}
\hypersetup{bookmarksnumbered=true}
\hypersetup{pdfstartview=FitH} 
\hypersetup{colorlinks=false}
\hypersetup{citebordercolor={.5 1 .5}}
\hypersetup{urlbordercolor={.5 1 1}}
\hypersetup{linkbordercolor={1 .7 .7}}

\allowdisplaybreaks[3]

\usepackage[font=small,labelfont=bf,width=0.85\textwidth]{caption}

\else

\documentclass[amsmath,amssymb,aps,letterpaper,prl,showpacs,preprint
,
]{revtex4}

\usepackage{graphicx}
\usepackage{color}
\fi

\makeatletter \let\@keywords\@empty \let\@subject\@empty
\providecommand{\keywords}[1]{\gdef\@keywords{#1}}
\providecommand{\subject}[1]{\gdef\@subject{#1}}
\def\thetitle{\@title}
\def\theauthor{\@author}
\def\thesubject{\@subject}
\def\thedate{\@date}
\def\thekeywords{\@keywords}
\makeatother
\AtBeginDocument{%
\hypersetup{pdftitle={\thetitle}}%
\hypersetup{pdfauthor={\theauthor}}%
\hypersetup{pdfsubject={\thesubject}}%
\hypersetup{pdfkeywords={\thekeywords}}%
}

\usepackage[bookmarks=true,hyperfigures=true]{hyperref}
\usepackage{fixmath}
\usepackage{mathtools}

\providecommand{\href}[2]{#2}

\let\oldbfseries=\bfseries
\let\oldmdseries=\mdseries
\let\oldnormalfont=\normalfont
\renewcommand{\bfseries}{\oldbfseries\boldmath}
\renewcommand{\mdseries}{\oldmdseries\unboldmath}
\renewcommand{\normalfont}{\oldnormalfont\unboldmath}

\makeatletter
\newlength{\apb@width}
\newcommand{\autoparbox}[2][c]{\settowidth{\apb@width}{#2}\parbox[#1]{\apb@width}{#2}}
\newcommand{\includegraphicsbox}[2][]{\autoparbox{\includegraphics[#1]{#2}}}
\makeatother

\newcommand{\nn}{\nonumber}

\newcommand{\be}{\begin{equation}}
\newcommand{\ee}{\end{equation}}
\newcommand{\ba}{\begin{eqnarray}}
\newcommand{\ea}{\end{eqnarray}}
\newcommand{\R}{\mathbb{R}}                                

\ifx\genfrac\sdflkaj\else\fi
\newcommand{\sfrac}[2]{{\textstyle\frac{#1}{#2}}}
\newcommand{\half}{\sfrac{1}{2}}

\newcommand \widebar [1] {\overline{#1}}

\newcommand{\ft}[2]{{\textstyle\frac{#1}{#2}}}
\newcommand{\cL}{\mathcal{L}}

\newcommand{\cO}{\mathcal{O}}

\def\l<{\langle}\def\r>{\rangle}

\makeatletter
\newcommand{\namedref}[2]{\hyperref[#2]{#1~\ref*{#2}}}
\newcommand{\secref}{\@ifstar{\namedref{Section}}{\namedref{sec.}}}
\newcommand{\subsecref}{\@ifstar{\namedref{Subsection}}{\namedref{subsec.}}}
\newcommand{\appref}{\@ifstar{\namedref{Appendix}}{\namedref{app.}}}
\newcommand{\tabref}{\@ifstar{\namedref{Table}}{\namedref{tab.}}}
\newcommand{\figref}{\@ifstar{\namedref{Figure}}{\namedref{fig.}}}
\makeatother

\newcommand{\abs}[1]{|#1|}

\newcommand{\eqn}[1]{(\ref{#1})}

\title{Quantum Gravitational Contributions to the Standard Model
Effective Potential and Vacuum Stability}
\begin{document}
%
\ifarxiv
\pdfbookmark[1]{Title Page}{title}

%

\thispagestyle{empty}

\begingroup\raggedleft\footnotesize\ttfamily
HU-EP-15/07\\
\vspace{15mm}
\endgroup

\begin{center}

\begingroup\Large\bfseries\thetitle\par\endgroup
\vspace{15mm}

\begingroup\large\scshape
Florian Loebbert and
Jan Plefka
\par
\endgroup

\vspace{5mm}

\begingroup\itshape
Institut f\"ur Physik and IRIS Adlershof,\\ 
Humboldt-Universit\"at zu Berlin, \\
Zum Gro{\ss}en Windkanal 6, D-12489 Berlin. Germany
\vspace{3mm}

\par\endgroup

{\ttfamily
\href{mailto:loebbert@physik.hu-berlin.de}{loebbert@physik.hu-berlin.de}\\
\href{mailto:plefka@physik.hu-berlin.de}{plefka@physik.hu-berlin.de}
}
\vspace{10mm}

\textbf{Abstract}

\bigskip

\begin{minipage}{13cm}

We compute the quantum gravitational contributions to the standard model effective potential and analyze their effects on the Higgs vacuum stability in the framework of effective field theory.
Einstein gravity necessarily implies the existence of higher dimension $\phi^{6}$ and $\phi^{8}$ operators with novel couplings $\eta_{1/2}$ in the Higgs sector.
  The beta functions of these couplings are established and the impact of the gravity induced contributions on electroweak vacuum stability is studied. We find that the true minimum of the standard model effective potential now lies below the Planck scale for almost the entire parameter space ($\eta_{1/2}(m_\text{t})> 0.01$). In addition quantum gravity is shown to contribute to the minimal value of the standard model NLO effective potential at the percent level.
The quantum gravity induced contributions yield a metastable vacuum
for a large fraction of the parameter space in the flowing couplings $\eta_{1/2}$.

\end{minipage}

\end{center}

\setcounter{page}{0}

\newpage
\else


\title{Quantum Gravitational Contributions to the Standard Model \\
Effective Potential and Vacuum Stability}

\preprint{HU-EP-15/07}

\author{Florian Loebbert}%
 \email{loebbert@physik.hu-berlin.de}
\author{Jan Plefka}%
 \email{plefka@physik.hu-berlin.de}
\affiliation{%
Institut f\"ur Physik and IRIS Adlershof,
Humboldt-Universit\"at zu Berlin, \\
Zum Gro{\ss}en Windkanal 6, D-12489 Berlin, Germany
}%


\begin{abstract}
We compute the quantum gravitational contributions to the standard model effective potential and analyze their effects on the Higgs vacuum stability in the framework of effective field theory.
Einstein gravity necessarily implies the existence of higher dimension $\phi^{6}$ and $\phi^{8}$ operators with novel couplings $\eta_{1/2}$ in the Higgs sector.
  The beta functions of these couplings are established and the impact of the gravity induced contributions on electroweak vacuum stability is studied. We find that the true minimum of the standard model effective potential now lies below the Planck scale for almost the entire parameter space ($\eta_{1/2}(m_\text{t})> 0.01$). In addition quantum gravity is shown to contribute to the minimal value of the standard model NLO effective potential at the percent level.
The quantum gravity induced contributions yield a metastable vacuum
for a large fraction of the parameter space in the flowing couplings $\eta_{1/2}$.
\end{abstract}

\pacs{
11.10.Hi,
14.70.Kv,
14.80.Bn,
04.60.Bc \hfill HU-EP-15/07
}


\maketitle

\fi

A central outcome of the recent Higgs boson discovery \cite{Aad:2012tfa,Chatrchyan:2012ufa}
and the absence of new physics signals at the LHC 
is that the standard model (SM) as a quantum field theory stays perturbatively self-consistent
all the way up to the Planck scale $M_{\text{Pl}}$ \cite{Degrassi:2012ry,Buttazzo:2013uya}. 
From this perspective the conservative scenario of having no beyond-SM-physics up to 
$M_{\text{Pl}}$---except for gravity---is viable and has attracted attention
\cite{Froggatt:1995rt,Meissner:2006zh,Shaposhnikov:2007nj,Shaposhnikov:2009pv,Holthausen:2011aa,Bezrukov:2014ipa}.%
\footnote{Of course neutrino masses, dark matter and baryogenesis still require
some (mild) extensions of the SM.}
Moreover, the measured values for the Higgs pole mass $m_\text{H}$ and the top mass $m_\text{t}$
have an intriguing consequence for the question of stability of the 
Higgs vacuum: The SM lies close to the border of absolute electroweak vacuum 
stability and metastability.
Vacuum stability is usually studied by determining the renormalization group improved
effective Higgs potential $V(\phi)$: 
If $V(\phi)$ develops a negative minimum below the value $V_\text{ew}$ of the electroweak minimum, 
the SM becomes
unstable; if the inverse decay rate for tunneling from the false electroweak minimum at 
$\phi=\phi_\text{ew}$ to the second (true) minimum at $\phi_{\text{min}}$ is larger than the lifetime of our universe, then the SM is said to be 
metastable. 
The value at the minimum $V_\text{min}=V(\phi_\text{min})<V_\text{ew}$ is very sensitive to
the value of $m_\text{H}$ and $m_\text{t}$. 
State of the art high precision perturbative calculations \cite{Degrassi:2012ry,Buttazzo:2013uya} 
using the latest experimental data 
indicate that the SM has a negative metastable minimum. However, it lies
deep within the Planck regime: The gauge invariant value of $V_{\text{min}}$ at 
next-to-leading order (NLO) precision
is $(-V_{\text{min}}^{\text{NLO}})^{1/4}\sim 10^{10}\, M_{\text{Pl}}$
\cite{Andreassen:2014gha}. While it is fascinating that the SM may be extrapolated 
to such high energy scales, it is obvious that quantum gravitational effects cannot be ignored
any longer in these regimes---even for the conservative no-new-physics scenario. 
In consequence, the celebrated statement of metastability of the SM based on the above value for 
$V_{\text{min}}$ is spurious. 
\begin{figure}[t]
\begin{center}
\includegraphics[width=8.7cm]{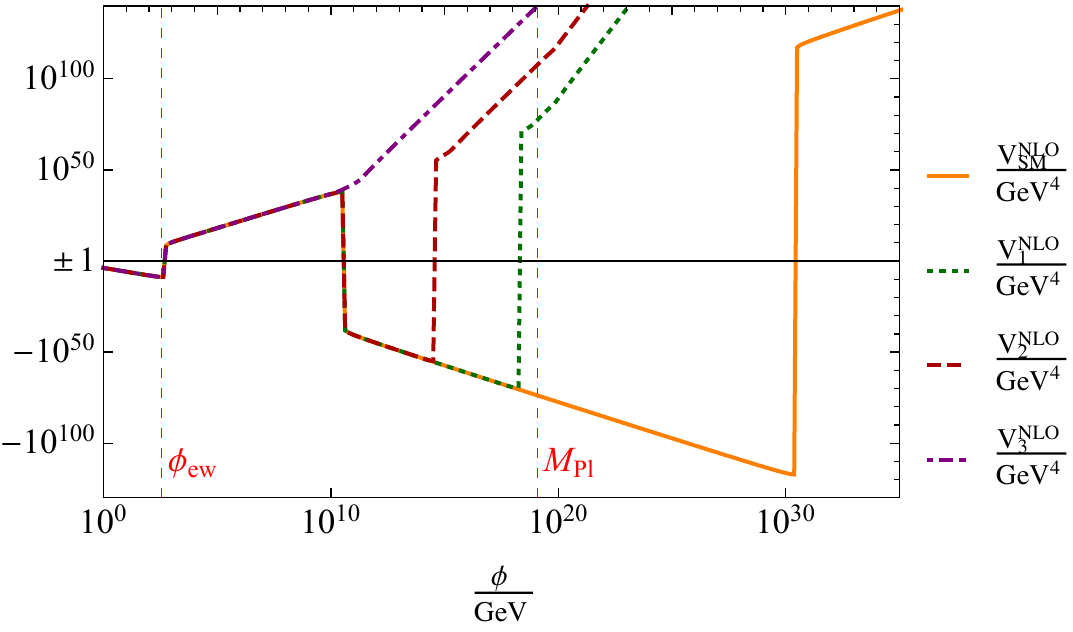}
\end{center}
\caption{
Double logarithmic plot of the NLO effective potential with and without gravitational corrections for sample initial values: (1)~green/dotted: $\eta_1(m_\text{t})=0.5$, $\eta_2(m_\text{t})=0.3$ with minimal value $V_{1,\text{min}}=-4.9\times 10^{69}~\text{GeV}^4$; (2)~red/dashed: $\eta_1(m_\text{t})=10^7$, $\eta_2(m_\text{t})=10^9$  with $V_{2,\text{min}}=-4.6\times10^{54}~\text{GeV}^4$;  (3)~purple/dashdotted: $\eta_1(m_\text{t})=10^{15}$, $\eta_2(m_\text{t})=10^{20}$ where the minimum has disappeared. (SM)~orange: 
Without gravity, we find $V_{\text{SM},\text{min}}=-4.0\times10^{110}~\text{GeV}^4$. 
}
\label{fig:teaser}
\end{figure}

This is the motivation for the present analysis. We adopt the conservative viewpoint
of having no new physics up to $M_{\text{Pl}}$ and study the quantum gravity contributions
to the SM effective potential. This is done by treating Einstein gravity as an effective 
quantum field theory \cite{Donoghue:1994dn}.
For scales below $M_{\text{Pl}}=1.22\times 10^{19}~\text{GeV}$, the SM coupled
to quantum gravity is perturbatively well defined albeit of
limited predictability due to the necessity of including higher dimensional operators as counterterms at every loop order. 
In fact, the main impact of the gravitational contributions to the Higgs effective potential is that 
non-renormalizability induces higher dimension counter terms of the form
$\frac{\eta_{1}}{M_{\text{Pl}}^{2}}\phi^{6}$ and $\frac{\eta_{2}}{M_{\text{Pl}}^{4}}\phi^{8}$ 
into the effective field theory, with a priory undetermined couplings $\eta_{1}$ and
$\eta_{2}$. The addition of such higher dimension operators to the tree-level potential 
has been studied in \cite{Greenwood:2008qp,Branchina:2013jra,Andreassen:2014gha} as a means to parametrize
potential new physics effects on top of the SM. 
Here we show that these terms are necessarily present due to the existence of gravity. Even if one chooses the new couplings to be absent at low scales, they are turned on at high scales by the quantum gravity contributions to the renormalization group equations (RGE).
The higher dimensional
counter terms have a profound effect: For generic positive values of 
$\eta_{1/2}(m_\text{t})$ the true minimum of the SM effective potential is pushed to sub-Planckian scales.
It is generically metastable or even stable for a large range of initial values of 
$\eta_{1/2}$ at the scale $m_\text{t}$. See Figure~\ref{fig:teaser} for generic configurations of $V(\phi)$ with
and without gravitational contributions.
On top, quantum gravity effects contribute to the effective potential at NLO with orders of magnitude at the percent level. 

\paragraph{Standard model coupled to Einstein gravity.} 
We consider the standard model coupled to gravity with a cosmological constant $\Lambda$:%
\footnote{Note that we could also include an additional non-minimally curvature-coupled scalar term  $2 \, \sqrt{-g}\, R\,  \rho\, |H|^2 $ into the Lagrangian. However, the freedom of Higgs-field redefinitions and Weyl rescalings may be used to set
$\rho=0$ in \eqn{model} \cite{Isidori:2007vm} at the cost of introducing an additional kinetic term. Here we set $\rho=0$ from the start.}
\begin{align}\label{model}
\cL^{\text{SM}}_{\text{grav}} = \frac{2}{\kappa^{2}} \, \sqrt{-g}\, R
 + \cL_{\text{GB}} + 
\cL_{\text{F}}
 + \sqrt{-g}\, \Bigl 
(\Lambda+g^{\mu\nu}\, \partial_{\mu}H
\,\partial_{\nu}H^{\dagger} + m^2 \, |H|^2 - \lambda\, |H|^{4} \Bigr).
\end{align}
Here $\kappa$ denotes the dimensionful gravitational coupling constant related 
to Newton's constant
$
\kappa^{2} = 32\pi G= M_{\text{Pl}}^{-2}
$.
As we are
interested in the quantum gravitational contributions to the Higgs effective potential
at the one-loop order, it is sufficient to study the Higgs-gravity sector of the above model
since the gauge bosons in
$\cL_{\text{GB}}$ and the matter fermions in $\cL_{\text{F}}$ decouple at this leading perturbative order. 
Moreover, due to the
non-renormalizability of gravity we will be forced to include higher dimension scalar
operators as counterterms
\be
\cL_{CT}= \sqrt{-g}\, \Bigl 
(-\eta_{1}\, \kappa^{2}\, |H|^{6} -\eta_{2}\, \kappa^{4}\, |H|^{8} \Bigr) \, ,
\ee
carrying their own dimensionless bare couplings $\eta_{1}$ and $\eta_{2}$.
We seek the one-loop corrections to the SM effective potential $V(\phi)$, where
we expand the Higgs doublet about the constant real background field $ \phi$
with 
$H=\frac{1}{\sqrt 2}\left (\begin{matrix}
\varphi_{1}+i\psi_{1} \cr \phi + \varphi_{2}+i\psi_{2} \end{matrix}\right )$ and
$\{\varphi_{i},\psi_{i}\}\in\R$. The tree-level Higgs potential is then given by
\be
V^{\text{tree}}(\phi)=-\frac{m^{2}}{2}\phi^{2} + \frac{\lambda}{4}\phi^{4}
+ \frac{\eta_{1}}{8}\kappa^{2}\phi^{6} + \frac{\eta_{2}}{16}\kappa^{4}\phi^{8}\, .
\ee
In the gravitational sector we work in de~Donder gauge expanding the metric
field as $g_{\mu\nu}=\eta_{\mu\nu}+ \kappa\, h_{\mu\nu}$. This yields the following
standard  expansions of
the Einstein--Hilbert and de Donder gauge fixing terms $\cL_\text{g.f.}$:
\begin{align}
\frac{2}{\kappa^{2}}\,\sqrt{-g} &R + \cL_\text{g.f.}= \ft{1}{2} h_{\alpha\beta}\, P^{\alpha\beta;\gamma\delta}\,\partial^{2}h_{\gamma\delta} 
+ \cO(h^{3}),\\
\sqrt{-g}&= 1 + \frac \kappa 2 \, h_{\alpha\beta}\eta^{\alpha\beta} + \frac{\kappa^{2}}{4}\,
h_{\alpha\beta}\, P^{\alpha\beta;\gamma\delta}\, h_{\gamma\delta}
+ \cO(h^{3}), \nn\\
P^{\alpha\beta;\gamma\delta}&= \ft 12
( \eta^{\alpha\beta}\, \eta^{\gamma\delta} - \eta^{\alpha\gamma}\, \eta^{\delta\beta}
- \eta^{\alpha\delta}\, \eta^{\gamma\beta})\nn \,.
\end{align}
For the quadratic fluctuations of the graviton and Higgs field in our model described by \eqn{model} 
we then have
\begin{align}
\cL_{h_{\mu\nu},\text{H}}^{\text{quad}} =&
-\ft 1 2 h_{\alpha\beta}\left [P^{\alpha\beta;\gamma\delta} \big(-\partial^{2}+m_{A}^{2}\big)\right ] h_{\gamma\delta}  
- h_{\alpha\beta}\left [ \eta^{\alpha\beta}\, m^{2}_{B} \right ] \varphi_{2} \\\nn
&- \ft 1 2\varphi_{2} \left [ 
\partial^{2}+m^{2}_{C}\right ] \varphi_{2}
-\;\sum_{\mathclap{\Phi_{I}=\{\varphi_{1}, \psi_{1},\psi_{2}\}}}\;
\ft 1 2 
\Phi_{I}\,[\partial^{2}+m_{D}^{2} ] \, \Phi_{I}\nn
\, ,
\end{align}
with the effective masses
\begin{align}
m_{A}^{2}&= \frac{\kappa^{2}}{4} \big(-m^{2}\phi^{2}+\ft{1}{2} \lambda\phi^{4}
+\ft{1}{4}\kappa^{2}\eta_{1}\phi^{6} +\ft{1}{8}\kappa^{4}\eta_{2}\phi^{8} +2\Lambda\big),\nn \\
m_{B}^{2} &= \frac{\kappa}{2}\, \big(-m^{2} \phi  +\lambda \phi^{3} +\ft{3}{4}
\kappa^{2}\eta_{1}\phi^{5} + \ft{1}{2}
\kappa^{4}\eta_{2}\phi^{7} \big), \nn \\
m^{2}_{C} &= -m^{2} + 3\lambda \phi^{2} + \ft{15}{4} \kappa^{2}\eta_{1}\phi^{4}
+\ft{7}{2}\kappa^{4}\eta_{2}\phi^{6},\nn \\
m^{2}_{D} &= -m^{2} + \lambda \phi^{2} + \ft{3}{4} \kappa^{2}\eta_{1}\phi^{4}
+\ft{1}{2}\kappa^{4}\eta_{2}\phi^{6} \, .
\end{align}
We see that the graviton obtains a small mass $m_A$ generated by the Higgs field. 
While beyond the scope of this letter, it would be important to further analyze the
SM coupled to gravity in
the light of different Higgs mechanisms for the graviton discussed in e.g.\ \cite{'tHooft:2007bf,Chamseddine:2010ub}.

\begin{figure}[t]
\begin{center}
$\lambda^2$\includegraphicsbox[width=1.8cm]{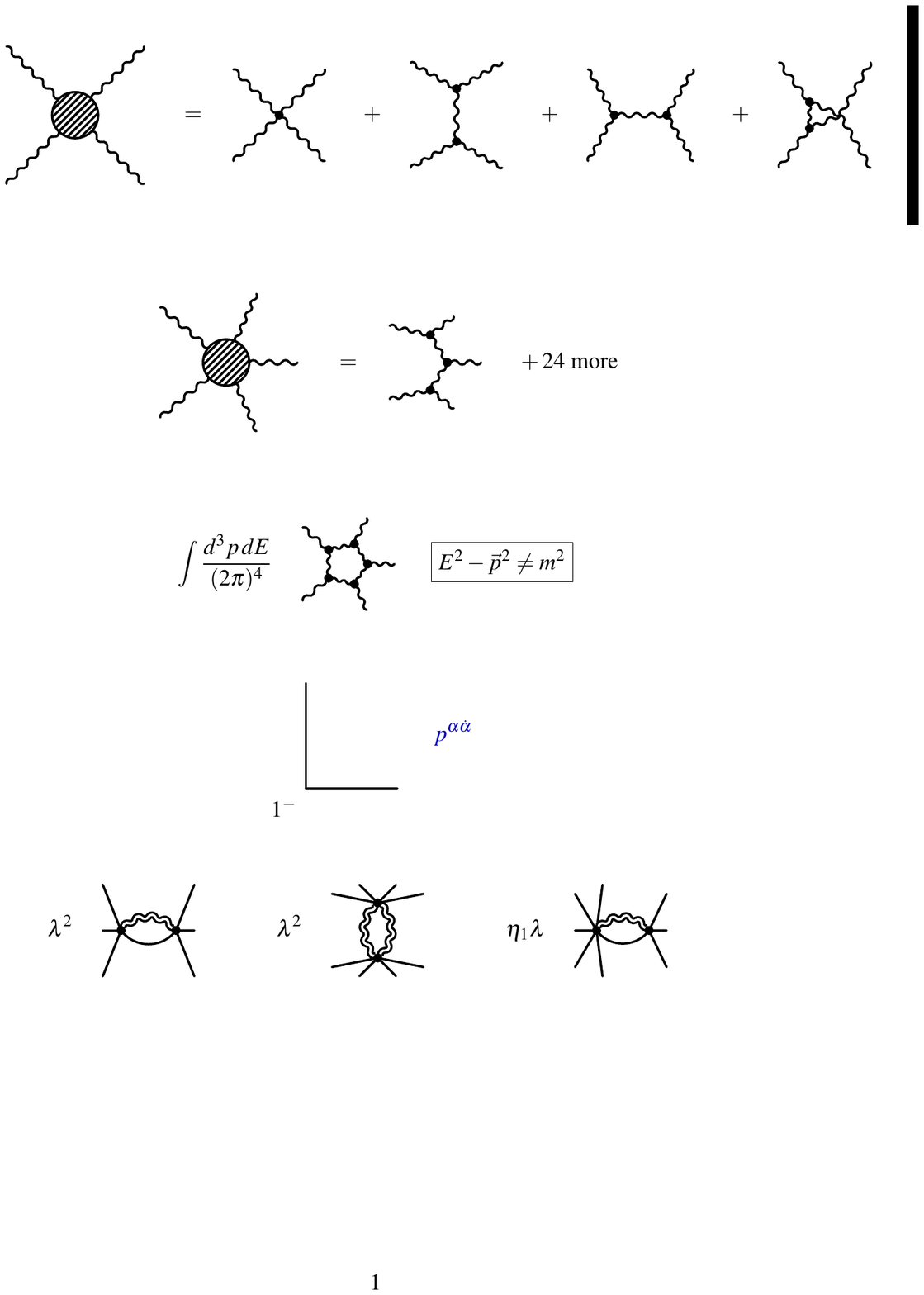}
\qquad
$\lambda^2$\includegraphicsbox[width=1.8cm]{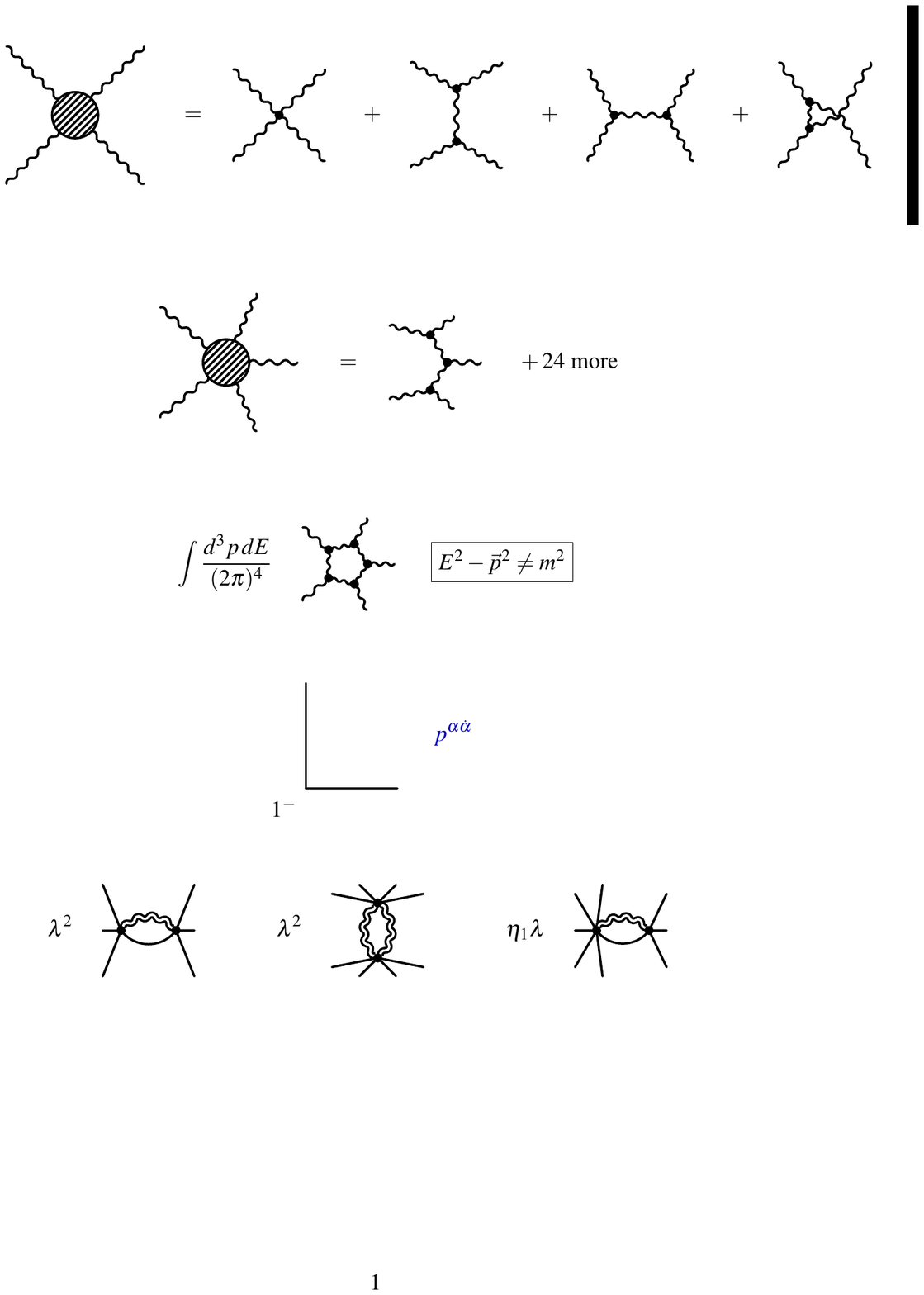}
\qquad
$\eta_1\lambda$\includegraphicsbox[width=1.8cm]{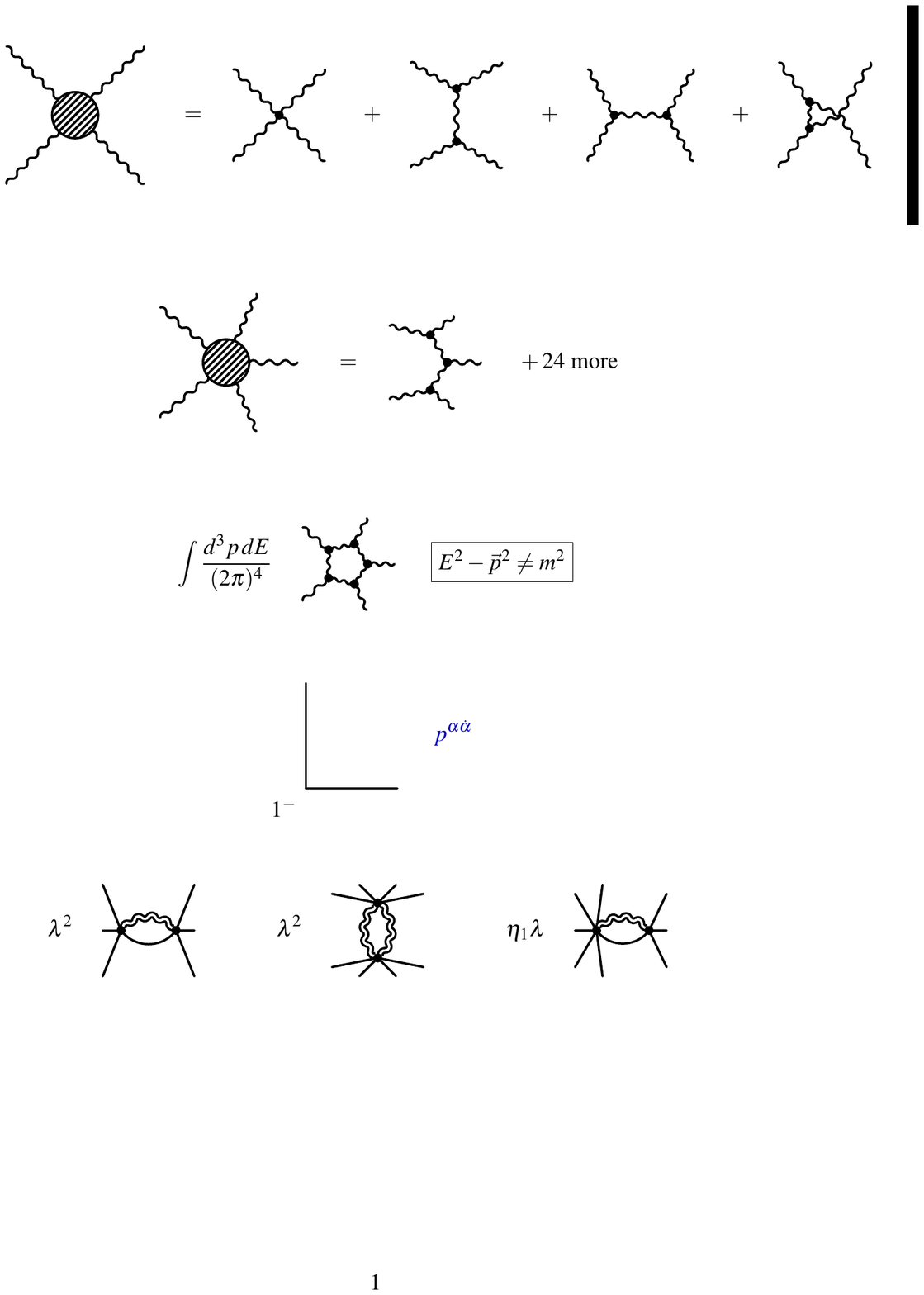}
\caption{
Diagrams of gravitational processes contributing to the $\beta$-functions of $\eta_{1}$
and $\eta_{2}$ in \eqn{betaeta1}. Plain lines represent the Higgs field,
wiggly lines the gravitons.}
\label{diagrams}
\end{center}
\end{figure}

Here we proceed by writing $\cL_{h_{\mu\nu},\text{H}}^{\text{quad}}= - \half V_{I} M^{IJ} V_{J}$ with the
collective quantum field $V_{I}=(h_{\mu\nu}, \varphi_{i},\psi_{i})$. We perform the path integral at 1-loop order to find the
gravitational contributions $V^{(\text{1-loop})}_{\text{grav}}= \Delta V[\phi] -\Delta V[\phi]\bigr |_{\kappa\to0}$ to the
effective 1-loop Higgs potential, where%
\footnote{We have $\det M = 
{64} 
(-\partial^{2}+m_{A}^{2})^{9}\, 
[ (-\partial^{2}+m_{A}^{2})(\partial^{2}+m_{C}^{2})-4m_{B}^{4}] (\partial^{2}+ m^{2}_{D})^{3}$.
}
\begin{align}\label{DVgrav}
\Delta V[\phi] =
-\frac{i}{2} \bar\mu^{4-d}\int \frac{d^{d}p}{(2\pi)^{d}}\, \Bigl ( 
9\ln(p^{2}+m_{A}^{2}) 
+ 3 \ln(p^{2}-m_{D}^{2}) 
+\ln [(p^{2}+m_{A}^{2})(p^{2}-m^{2}_{C})-4m_{B}^{4}] 
\Bigr ) \, .
\end{align}
The relevant dimensionally regularized integral reads
\begin{align}
\label{drintegral}
-\frac{i}{2}\, \bar\mu^{4-2\epsilon}\, \int \frac{d^{4-2\epsilon}p}{(2\pi)^{4-2\epsilon}}
\ln (p^{2}-m^{2}) =   - \frac{m^{4}}{64\pi^{2}} \frac{1}{\epsilon} 
+ \frac{m^{4}}{64\pi^{2}}\,  \left [ \ln \bigg(\frac{m^{2}}{\mu^{2}}\bigg) - \frac{3}{2}\right ],
\end{align}
with $\mu^{2}=4\pi\bar\mu^{2} e^{-\gamma_{E}}$%
\footnote{We stress the use of dimensional regularization here, which only sees logarithmic divergences. A 
(naive) cutoff regularization $|p|<\Lambda_{\text{UV}}$ of the integral would also induce a $\Lambda_{\text{UV}}^{2} m^{2}$
power divergence in \eqn{drintegral}. However, this term does not contribute to the $\beta$-functions
in perturbatively renormalized effective field theory. Nevertheless, in the framework of the functional 
renormalization group approach this term contributes, c.f.~\cite{Eichhorn:2015kea}.}. 
The pole in ${\epsilon}$ 
yields a renormalization of the scalar field couplings $m,\lambda, \eta_{1}$ and $\eta_{2}$.
Proceeding in the $\widebar{\text{MS}}$ scheme gives the following gravitational contribution to
the renormalized effective potential of the standard model:
\begin{align}
 \Delta V&[\phi] =
\frac{9}{64\pi ^{2}}\, m_{A}^{4}\, 
\left (\ln \frac{m_{A}^{2}}{\mu^{2}}-\frac{3}{2} \right) 
+ 
3 \frac{m_{D}^{4}}{64\pi ^{2}}\, \left (\ln \frac{m_{D}^{2}}{\mu^{2}}-\frac{3}{2}\right) + 
\sum_{i=\pm}
\frac{C_{i}^{2}}{64\pi ^{2}}\, \left (\ln \frac{C_{i}}{\mu^{2}} -\frac{3}{2}\right) 
\, ,
\label{Vefftot}
\end{align}
where for conciseness we have defined%
\footnote{Note that the term under the square root is always positive at $\phi=\phi_\text{min}$ on the parameter space of $\eta_{1,2}$.}
\be
C_{\pm}= \frac{1}{2}\left ( 
m^{2}_{C} -m^{2}_{A}\pm \sqrt{ 
\left (m^{2}_{C} + m_{A}^{2} \right )^{2} - 16 m^{4}_{B}}
\right ) \, .
\ee
\paragraph{The $\beta$-functions.} 
Adding the counterterms necessary to absorb the $1/\epsilon$-pole terms to the bare Lagrangian yields the renormalized Lagrangian. An equivalent statement is that the effective
potential obeys a renormalization group equation (RGE) 
\be 
 \Big(\mu \ft\partial{\partial \mu }+\sum\nolimits_{i}\beta_{i}\ft{\partial}{\partial \lambda_{i} }
-\gamma_{\phi}\,\phi\ft\partial{\partial \phi}\Big)\, V_{\text{eff}}(\phi)=0\, ,
\ee
where $\beta_{\lambda_{i}}=\ft{d\lambda_{i}}{d\log \mu}$ are the $\beta$-functions of
the couplings $\lambda_{i}$ and $\gamma_{\phi}$ is the anomalous dimension of the 
SM Higgs field. From \eqn{Vefftot} one thus establishes the one-loop
$\beta$-functions of the novel couplings $\eta_{1/2}$,
which take the following form in the $\{\kappa^{2}m^{2},\kappa^{4}\Lambda\}\to 0$ limit:
\begin{align}
\beta^{(1)}_{\eta_{1}} &=
6\eta_{1}\gamma^{(1)}_{\phi}+ \ft{1}{16\pi^{2}} \Bigl [ 108\, \lambda\, 
\eta_{1}
- 8\lambda^{2}\Bigr ],  \label{betaeta1}\\
\beta_{\eta_{2}}^{(1)}&=
  8\eta_{2}\gamma^{(1)}_{\phi}+\ft{1}{16\pi^{2}}\Bigl [
 192 \lambda  \, \eta _2+126 \eta _1^2 \label{betaeta2} 
+ \ft 5 4 \lambda^{2}   -24 \eta _1 \lambda 
\Bigr ]\, , \nn \\
\gamma^{(1)}_{\phi}&= \ft{1}{16\pi^{2}} \Bigl [3y_{t}^{2}-\ft 3 4 g_{1}^{2}-\ft 9 4 g_{2}^{2}
\Bigr ]\,  . \nn
\end{align}
The scaling dimension $\gamma_{\phi}$ contains the top-Yukawa and 
electroweak coupling constants $y_{t}, g_{1}, g_{2}$.
We stress that the $\lambda^{2}$ term in $\beta^{(1)}_{\eta_{1}}$
 as well as the $\lambda^{2}$ and $\eta_{1}\lambda$
terms in $\beta^{(1)}_{\eta_{2}}$
are quantum gravity induced contributions despite the fact that they are not proportional to 
$\kappa$, see Figure~\ref{diagrams}. 
The remaining non-gravitational term in
$\beta_{\eta_{1}}^{(1)}$ has been reported in \cite{Jenkins:2013zja},  the non-gravitational terms in 
$\beta_{\eta_{2}}^{(1)}$ 
were considered in \cite{Lalak:2014qua}, however we disagree with the numerical factor for the 
$\lambda\eta_{2}$ term given there. 
We hence see that even in the absence of higher dimensional operators ($\eta_{1}=
\eta_{2}=0$) at an initial scale, these terms will be created in the renormalization group flow.%
\footnote{We assume that contributions to the RG-flow from other higher dimensional operators in the effective field theory are consistently neglectable at this order of perturbation theory. It would be important to investigate this point in full detail, cf.\ e.g.\ \cite{Jenkins:2013zja,Burgess:2014lza}. }
Including the Higgs mass and cosmological constant in the analysis leads 
to quantum gravitational contributions of 
order $\kappa^{2}m^{2}$ and $\kappa^{4}\Lambda$
to all the $\beta$-functions including $\beta_{\lambda}$ and $\beta_{m}$. 
However, these terms are of order $10^{-16}$ or less and absolutely negligible. 
We therefore
set $m=\Lambda=0$ in the remaining analysis. This also puts the electroweak minimum to zero, i.e.\
$V_\text{ew}=0$.

\paragraph{Consistent perturbation theory.}

The renormalization group improved SM effective potential is traditionally written in the form
\begin{equation}
\label{Veff1}
V_\text{SM}(\phi)=\lambda_\text{eff}(\mu=\phi)\frac{\phi^4}{4},
\end{equation}
with the effective field-dependent coupling constant 
\begin{equation}
\lambda_\text{eff}(\phi)=e^{4\Gamma(\phi)}\big[\lambda(\mu)+\lambda_\text{eff}^{(1)}(\mu)+\lambda_\text{eff}^{(2)}(\mu)\big]\big|_{\mu=\phi},
\end{equation}
where $\Gamma(\phi)=\int_{m_{t}}^{\phi}\gamma_{\phi}(\mu)d\log\mu$.
The explicit expressions for the corrections to the effective coupling 
$\lambda_\text{eff}$ up to three loops are given in the appendix of \cite{Buttazzo:2013uya}. 
We follow a consistent use of perturbation theory   
along the lines of \cite{Coleman:1973jx,Andreassen:2014eha,Andreassen:2014gha}
assuming $\lambda\sim\hbar$. This is necessary in order  
for the tree-level $\lambda\phi^{4}$ term to receive non-negligible corrections by the one-loop contribution scaling as
$y_{t}^{4}\sim\hbar$. Perturbation theory in $\hbar$  is applicable, however,
it is not the usual loop expansion. 
The expansion of the tree, one- and two-loop contributions of \eqn{Veff1}  along these lines
yields the expansion up to next-to-leading order:
\begin{equation}
V_{\text{SM}}^\text{NLO}(\phi)=V_{\text{SM}}^{\text{LO}}(\phi)+V_{\text{SM}}^{(\text{NLO})}(\phi)\, ,
\end{equation}
where $V_{\text{SM}}^{\text{LO}}$ scales as $\hbar$ and $V_{\text{SM}}^{(\text{NLO})}$ as $\hbar^2$.

Let us now add the gravitational contributions to this picture. 
In analogy to the SM case above we write
\begin{equation}
\label{etaeff}
V_\text{grav}(\phi)=\eta_\text{eff}(\mu=\phi)\frac{\phi^4}{4},
\end{equation}
with the RG-improved effective coupling
\begin{align}
\eta_\text{eff}(\phi)=& \big[e^{6\Gamma(\phi)}\eta_1(\mu)\kappa^2 \phi^2+
e^{8\Gamma(\phi)}\eta_2(\mu)\kappa^4\phi^4 
 +  \eta^{(1)}_\text{eff}(\phi)\big]\big|_{\mu=\phi}\, .
\end{align}
Here $\eta^{(1)}_\text{eff}(\phi)= 4V^{(\text{1-loop})}_{\text{grav}}/\phi^{4}$
can be extracted from \eqn{Vefftot}.
To obtain a consistent perturbative expansion we assume
$\eta_1\kappa^{2}\phi^{2}\sim\hbar$ and $\eta_2\kappa^{4}\phi^{4}\sim\hbar$. 
Expanding $V_\text{grav}(\phi)$ in $\hbar$ then yields the gravitational correction to the SM
effective potential up to next-to-leading order $\hbar^{2}$:
\begin{equation}
V^{\text{NLO}}_\text{grav}(\phi)=V^{\text{LO}}_\text{grav}(\phi)+V^{(\text{NLO})}_\text{grav}(\phi).
\end{equation}
Both, $V_{\text{SM}}^{\text{NLO}}$ and $V^{\text{NLO}}_\text{grav}$ have nonzero imaginary parts and in the following we will restrict to the real part of the potential, referring to 
\cite{Weinberg:1987vp} 
for an interpretation of the imaginary contribution. 

\paragraph{Minimum of the effective potential.}

The full leading order (LO) potential reads%
\footnote{See \cite{Andreassen:2013hpa} for a good review on effective potentials in the context of the SM. }
\be
V^{\text{LO}}(\phi)=
V^{\text{tree}}(\phi)
 - \frac{3}{64} y_{t}^{4}\,\phi^{4}\, \log\frac{y_{t}^{2}\phi^{2}}{2\mu^{2}}
+(\text{$g_{1},g_{2}$ terms})\,,
\nn
\ee
where we suppress the numerically small gauge coupling terms for brevity. The 
position of the true minimum $\phi_{\text{min}}=\mu_{\text{min}}$ is then the second nontrivial solution of 
$\frac{d}{d\phi}V^{\text{LO}}=0$ with flowing couplings, guaranteeing gauge invariance
of the minimum, c.f.~\cite{Andreassen:2014eha,Andreassen:2014gha}.
At NLO for the full effective potential we then have
\be
V_{\text{min}}= V^{\text{LO}}(\phi_{\text{min}})+
V_{\text{SM}}^{(\text{NLO})}(\phi_{\text{min}})  +V^{(\text{NLO})}_\text{grav}(\phi_{\text{min}})\, .
\ee
\begin{figure}[t]
\begin{center}
\includegraphicsbox[scale=.7]{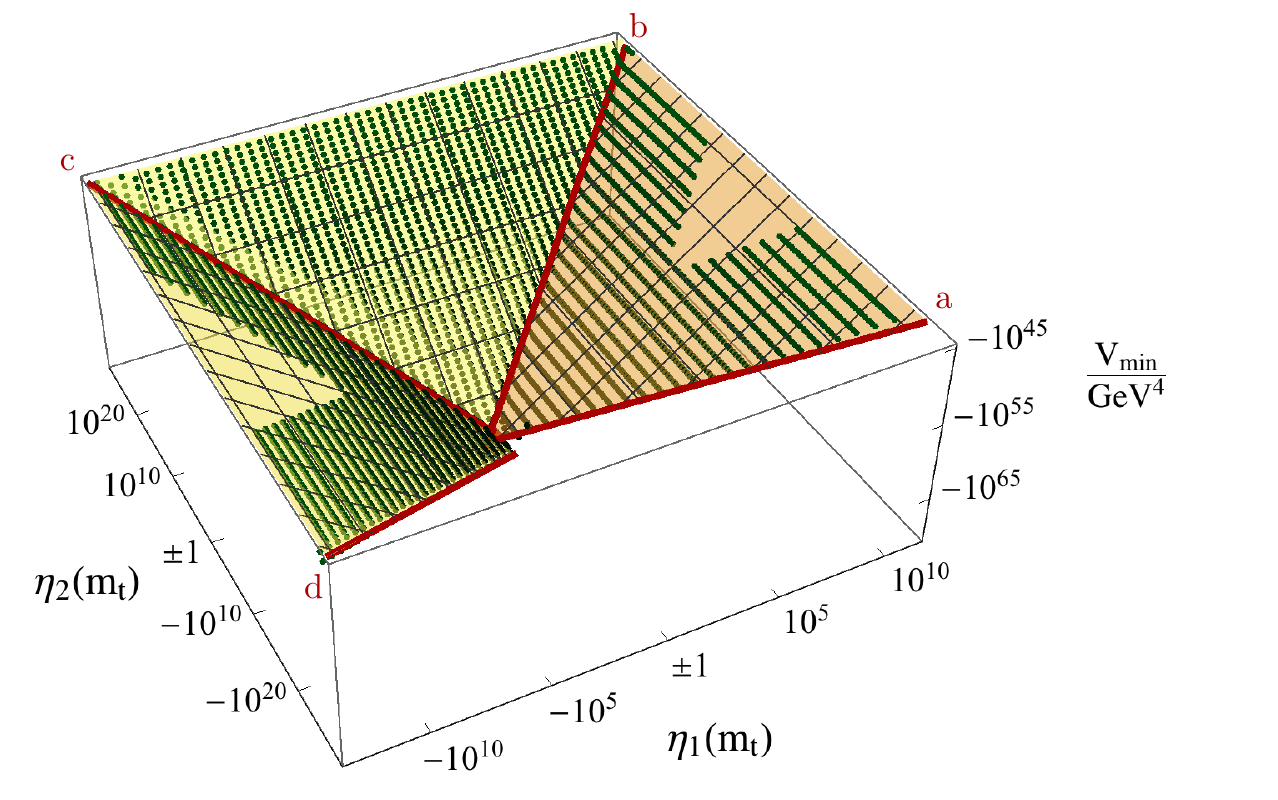}\\
\end{center}
\caption{
Plot of $V_\text{min}$ for different initial values $\eta_1(m_\text{t})$ and $\eta_2(m_\text{t})$ at the top mass scale (green dots). In the triangular regions the huge ratio $\sfrac{\eta_1(m_\text{t})}{\eta_2(m_\text{t})}$ yields a breakdown of the numerical solution to the RGEs such that no data points are provided. The three (yellow) planes fit the data to good accuracy. At the boundaries of the $\eta_{1/2}$-plane, the value at the minimum becomes positive (before the minimum disappears).}
\label{fig:valmin}
\end{figure}%
%
In Figure~\ref{fig:valmin} we plot the explicit data for $V_{\text{min}}$ as a function of the initial values $\eta_1(m_\text{t})$ and $\eta_2(m_\text{t})$ at the top mass scale. Here we use the initial conditions provided in \cite{Buttazzo:2013uya,Andreassen:2014gha}, in particular $m_\text{t}=173.34~\text{GeV}$ and $m_\text{H}=125.14~\text{GeV}$, the available higher loop RG
equations for the SM and our one-loop RGEs for $\eta_{1/2}$.
In large regions of the parameter space, the value at the minimum is well approximated by the three planes depicted in the double logarithmic plot in Figure~\ref{fig:valmin} (bounded by the four (red) lines):
\begin{align} 
V_\text{min}^{\text{ab}}&\simeq
-\sfrac{2}{\eta_1^2\kappa^4 10^8}, 
&
V_\text{min}^{\text{bc}}&\simeq
-\sfrac{1}{\eta_2\kappa^4 10^5}, 
&
V_\text{min}^{\text{cd}}&\simeq
-\sfrac{1}{\eta_1^2\kappa^4 10^5}. \nn
\end{align}
Here the lines bounding the planes are parametrized by
(a=d)~$\eta_2=-30\eta_1^2$, 
(b)~$\eta_2=500\eta_1^2$ for $\eta_1>0$,
(c)~$\eta_2=-\eta_1^2$ for $\eta_1<0$.
Similar values were reported in \cite{Andreassen:2014gha} for $\eta_2\equiv0$ and
non-flowing $\eta_{1}$. 
The minimum disappears approximately on the lines $\eta_1(m_\text{t})\simeq10^{14}$ and $\eta_2(m_\text{t})\simeq 10^{31}$ in the $\eta_{1/2}$-plane. The value $V_\text{min}$ turns positive shortly before, at around $\eta_1(m_\text{t})\simeq10^{12}$ and $\eta_2(m_\text{t})\simeq 10^{30}$. 
Such astronomically high values for $\eta_{1/2}$ should be read as measures of scales
$M$ when new physics arises in the sense of $M=M_{\text{Pl}}/\sqrt{\eta_{1}}$ and
$M=M_{\text{Pl}}/\eta_{2}^{1/4}$. In that sense the above
thresholds reflect the existence of an intriguing stability scale $M\sim 10^{10}$~GeV.
Finally, for positive $\eta_{1/2}(m_\text{t})\lesssim0.01$ the minimum lies beyond the Planck scale. 

Let us briefly discuss the order of magnitude of the NLO gravity contributions to the effective potential. The ratio $V_\text{grav}^\text{(NLO)}(\phi_\text{min})/V_\text{min}$ evaluated at the minimum ranges between zero and ten percent in large regions of the parameter space. At the boundaries, for large initial $\eta_1(m_\text{t})$ or $\eta_2(m_\text{t})$, the minimal value $V_\text{min}$ changes sign as indicated above and the relative gravitational contribution becomes large. 
\begin{figure}[t]
\begin{center}
\includegraphicsbox[width=5.9cm]{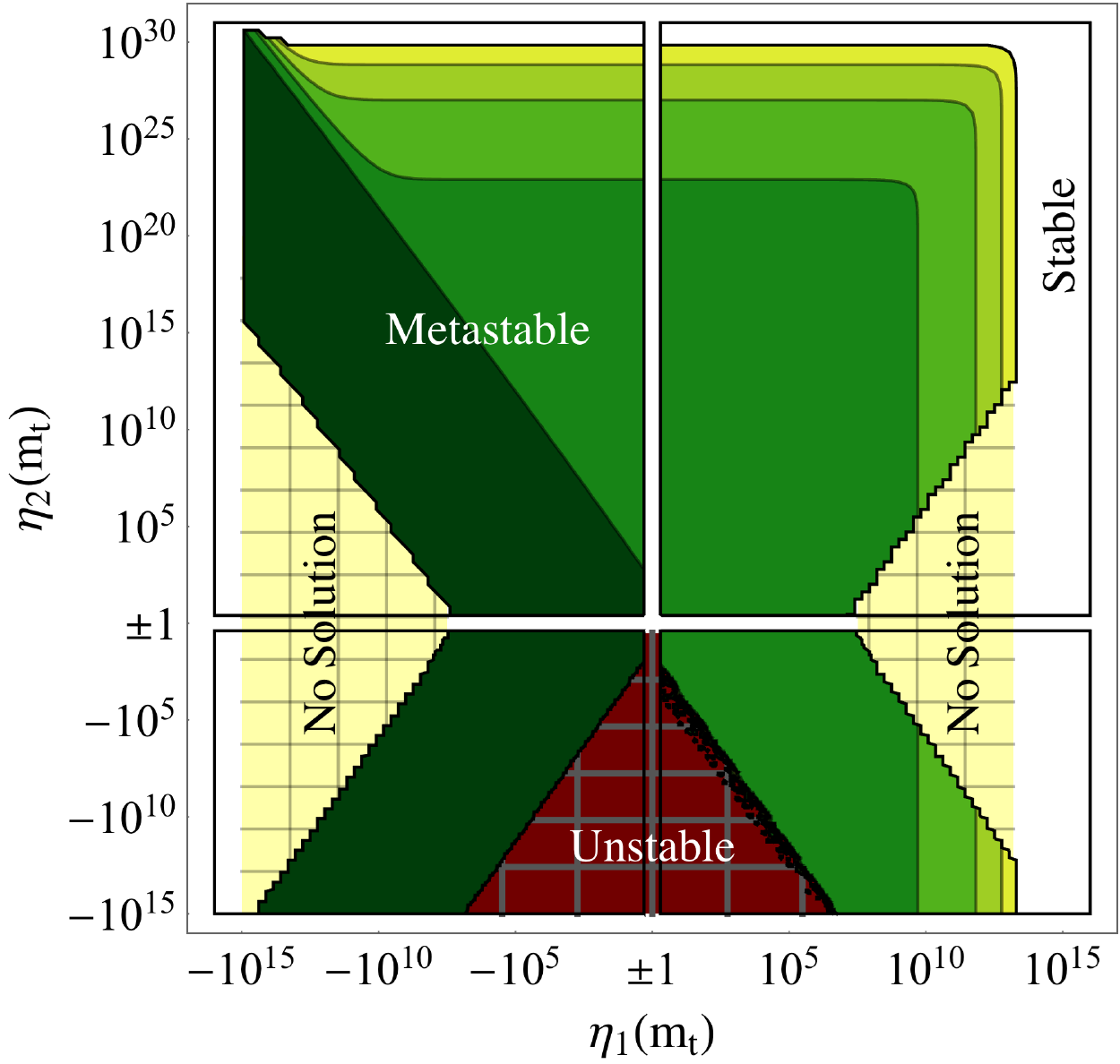}
\,\,
\begin{minipage}{2.5cm}
\includegraphicsbox[width=1.3cm]{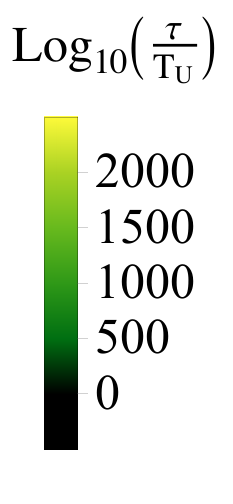}\hfill
\includegraphicsbox[width=2.5cm]{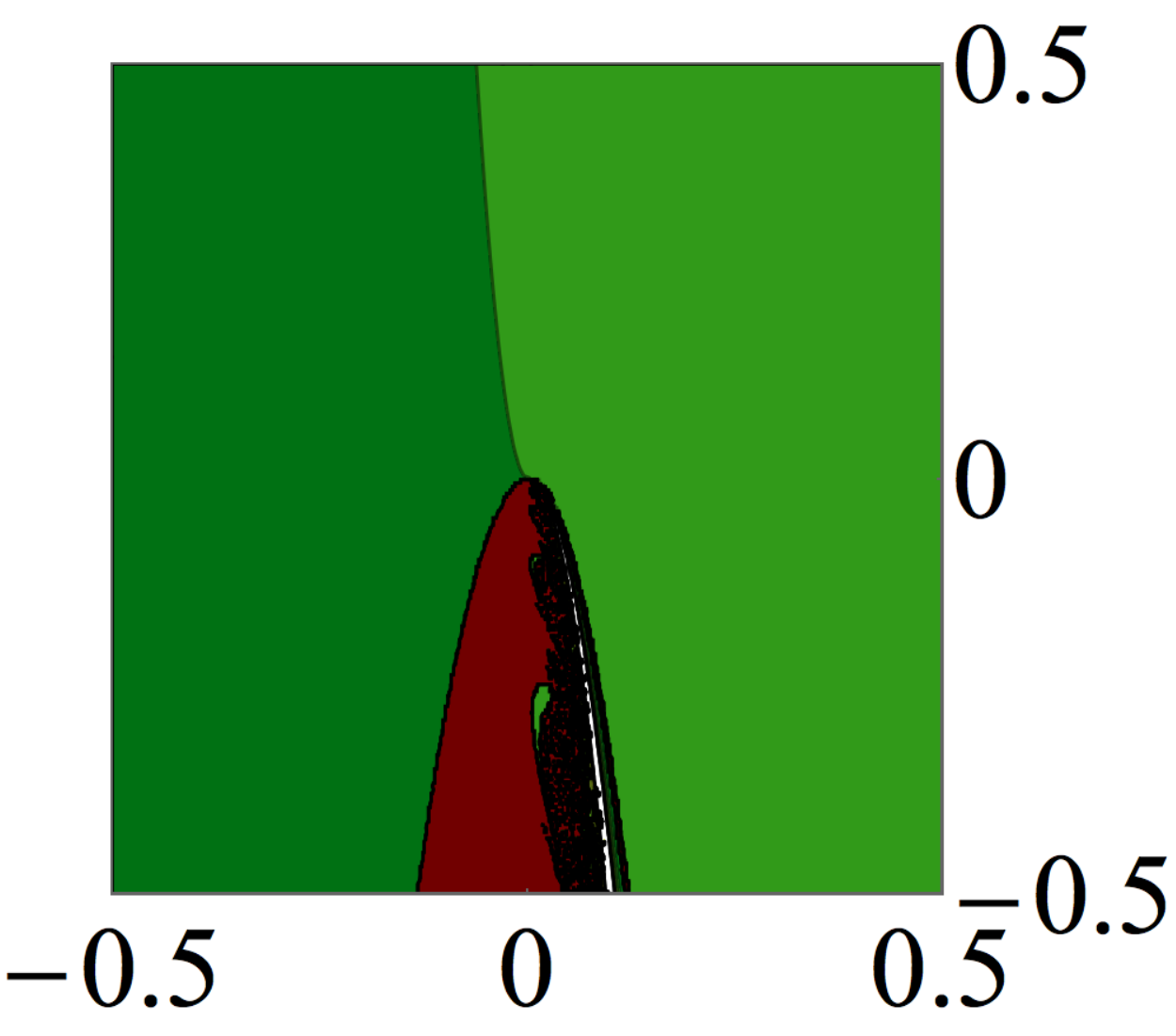}
\vspace{.8cm}
\end{minipage}
\end{center}
\caption{
Portrait of the Higgs vacuum lifetime as a function of 
$\eta_{1/2}$ at the top mass scale. The dark meshed (red) region denotes instability; the light meshed (yellow) region 
could not be explored due to numerical issues. The white areas at the boundaries of the plot denote the stability region. On the right hand side we magnify the region $\abs{\eta_{1/2}}\leq 0.5$.}
\label{fig:LifeTime}
\end{figure}

\paragraph{Impact on vacuum stability.} 
We have seen that for a huge range of the parameter space of $\eta_{1/2}$,
the SM coupled to gravity develops a negative minimum $V_{\text{min}}$ below the electroweak
one. The lifetime $\tau$ in units of the age of the universe $T_\text{U}$ of the false vacuum 
is usually estimated in the SM by \cite{Isidori:2001bm}
$
\frac{\tau}{T_\text{U}}= \min_{\mu}\frac{1}{(\mu\, T_\text{U})^4}\exp[\frac{8\pi^{2}}{3|\lambda(\mu)|}]
\, .
\label{LifetimeSM}
$
In the presence of the dimension six and eight counterterms this equation needs to be modified. Numerical
studies \cite{Branchina:2014usa,Branchina:2014rva} indicate that $\tau$ is still well
approximated by simply replacing $|\lambda(\mu)|$ in the exponent of the 
lifetime expression
by $|\lambda(\mu)+\eta_{\text{eff}}(\mu)|$. 
In Figure~\ref{fig:LifeTime} we show the lifetime of the false electroweak vacuum as a 
function of the initial values for $\eta_{1/2}$ at the top mass scale using this lifetime
approximation. We see that the SM coupled to gravity generically yields a highly metastable vacuum
for a huge part of the parameter space in $\eta_{1/2}$. Notably, the instability only occurs for negative
$\eta_{2}$, i.e. $\eta_{2}<-30\eta_{1}^{2}$. Moreover, we find that the ultrashort lifetimes
reported in \cite{Branchina:2014usa} are confined to a very small (black) region in $\eta_{1/2}(m_{t})$ parameter space once the RG-flow is taken into account.

Certainly the lifetime formula employed should be taken cautiously as it does not include curvature effects \cite{Isidori:2007vm,Burda:2015isa}.
In regions where $|V_{\text{min}}|^{1/4}$ comes close to $M_{\text{Pl}}$ the assumption of
a flat background metric turns inconsistent. In these regions a full analysis
expanding the metric around curved backgrounds should be employed (cf.\ e.g.\ \cite{Elizalde:1993ee} in this context). 

Beyond the instability region derived above the values of the
novel couplings $\eta_1$ and $\eta_{2}$ are not restricted by present observational
data, as so far only single Higgs interactions have been probed experimentally 
(see e.g.~\cite{Falkowski:2013dza} for a recent analysis)%
\footnote{Cf.\ \cite{Chu:2015nha} for a lattice approach towards restrictions on $\eta_1$.}.
Therefore also the exponentially
large values explored here are not excluded
and remain perturbative.

Finally, we note that the reported results on the values of $V_{\text{min}}$ and the lifetimes are sensitive to the values of $m_\text{t}$ and $m_\text{H}$ even at the level of a 2~GeV variation. Mapping this out is left for future work.



\paragraph{Acknowledgements.}
We thank P.~Galler, S.~Huber, K.~Meissner, H.~Nicolai, M.~Shaposhnikov and P.~Uwer for discussions.

\bibliography{bibliothek}

\begin{thebibliography}{10}
\bibitem{Shaposhnikov:2007nj}
\nbbstauthor{M.~Shaposhnikov},
\nbbsttitle{``{Is there a new physics between electroweak and Planck
  scales?}''},
\nbbsteprint{\arxivref{0708.3550}{arxiv:0708.3550}}.
\end{thebibliography}
\pdfbookmark[1]{\refname}{references}
\bibliographystyle{nb}

\end{document}